\documentclass[conference]{IEEEtran}

\ifCLASSINFOpdf
\else
\fi
\usepackage[bookmarks=false]{hyperref}
\usepackage{graphicx} % This is used to load the crest in the title page
\usepackage{amssymb}
\usepackage{amsmath}
\usepackage{hyperref}
\usepackage{array}
\usepackage[font=small,it]{caption}
\usepackage{epstopdf}

\hypersetup{
 colorlinks,
 citecolor=black,
 linkcolor=black,
 urlcolor=blue}

\begin{document}

\title{A reinforcement learning extension to the Almgren-Chriss framework for optimal trade execution}

\author{\IEEEauthorblockN{Dieter Hendricks}
\IEEEauthorblockA{School of Computational and\\Applied Mathematics\\
University of the Witwatersrand\\
Johannesburg, South Africa\\
Email: dieter.hendricks@students.wits.ac.za}
\and
\IEEEauthorblockN{Diane Wilcox}
\IEEEauthorblockA{School of Computational and\\Applied Mathematics\\
University of the Witwatersrand\\
Johannesburg, South Africa\\
Email: diane.wilcox@wits.ac.za}}

\maketitle

\begin{abstract}
%\boldmath
Reinforcement learning is explored as a candidate machine learning technique to enhance existing analytical solutions for optimal trade execution with elements from the market microstructure. Given a volume-to-trade, fixed time horizon and discrete trading periods, the aim is to adapt a given volume trajectory such that it is dynamic with respect to favourable/unfavourable conditions during realtime execution, thereby improving overall cost of trading. We consider the standard Almgren-Chriss model with linear price impact as a candidate base model. This model is popular amongst sell-side institutions as a basis for arrival price benchmark execution algorithms. By training a learning agent to modify a volume trajectory based on the market's prevailing spread and volume dynamics, we are able to improve post-trade implementation shortfall by up to 10.3\% on average compared to the base model, based on a sample of stocks and trade sizes in the South African equity market.
\end{abstract}

\IEEEpeerreviewmaketitle

\section{Introduction}
A critical problem faced by participants in investment markets is the so-called optimal liquidation problem, viz. how best to trade a given block of shares at minimal cost. Here, cost can be interpreted as in Perold's implementation shortfall (\cite{perold}), i.e. adverse deviations of actual transaction prices from an arrival price baseline when the investment decision is made. Alternatively, cost can be measured as a deviation from the market volume-weighted trading price (VWAP) over the trading period, effectively comparing the specific trader's performance to that of the average market trader. In each case, the primary problem faced by the trader/execution algorithm is the compromise between price impact and opportunity cost when executing an order. 

Price impact here refers to adverse price moves due to a large trade size absorbing liquidity supply at available levels in the order book (temporary price impact). As market participants begin to detect the total volume being traded, they may also adjust their bids/offers downward/upward to anticipate order matching (permanent price impact) \cite{holthausen}. To avoid price impact, traders may split a large order into smaller child orders over a longer period. However, there may be exogenous market forces which result in execution at adverse prices (opportunity cost). This behaviour of institutional investors was empirically demonstrated in \cite{chan}, where they observed that typical trades of large investment management firms are almost always broken up into smaller trades and executed over the course of a day or several days.

Several authors have studied the problem of optimal liquidation, with a strong bias towards stochastic dynamic programming solutions. See \cite{bertsimas}, \cite{huberman}, \cite{vayanos}, \cite{almgren} as examples. In this paper, we consider the application of a machine learning technique to the problem of optimal liquidation. Specifically we consider a case where the popular Almgren-Chriss closed-form solution for a trading trajectory (see \cite{almgren}) can be enhanced by exploiting microstructure attributes over the trading horizon using a reinforcement learning technique.

Reinforcement learning in this context is essentially a calibrated policy mapping states to optimal actions. Each state is a vector of observable attributes which describe the current configuration of the system. It proposes a simple, model-free mechanism for agents to learn how to act optimally in a controlled Markovian domain, where the quality of action chosen is successively improved for a given state \cite{dayan}. For the optimal liquidation problem, the algorithm examines the salient features of the current order book and current state of execution in order to decide which action (e.g. child order price or volume) to select to service the ultimate goal of minimising cost.

The first documented large-scale empirical application of reinforcement learning algorithms to the problem of optimised trade execution in modern financial markets was conducted by \cite{nevmyvaka}. They set up their problem as a minimisation of implementation shortfall for a buying/selling program over a fixed time horizon with discrete time periods. For actions, the agent could choose a price to repost a limit order for the remaining shares in each discrete period. State attributes included elapsed time, remaining inventory, current spread, immediate cost and signed volume. In their results, they found that their reinforcement learning algorithm improved the execution efficiency by 50\% or more over traditional submit-and-leave or market order policies.

Instead of a pure reinforcement learning solution to the problem, as in \cite{nevmyvaka}, we propose a hybrid approach which \textit{enhances} a given analytical solution with attributes from the market microstructure. Using the Almgren-Chriss (AC) model as a base, for a finite liquidation horizon with discrete trading periods, the algorithm determines the proportion of the AC-suggested trajectory to trade based on prevailing volume/spread attributes. One would expect, for example, that allowing the trajectory to be more aggressive when volumes are relatively high and spreads are tight may reduce the ultimate cost of the trade. In our implementation, a static volume trajectory is preserved for the duration of the trade, however the proportion traded is dynamic with respect to market dynamics. As in \cite{nevmyvaka}, a market order is executed at the end of the trade horizon for the remaining volume, to ensure complete liquidation. An important consideration in our analysis is the specification of the problem as a finite-horizon Markov Decision Process (MDP) and the consequences for optimal policy convergence of the reinforcement learning algorithm. In \cite{nevmyvaka}, they use an approximation in their framework to address this issue by incorporating elapsed time as a state attribute, however they do not explicitly discuss convergence. We will use the findings of \cite{garcia} in our model specification and demonstrate near-optimal policy convergence of the finite-horizon MDP problem.

The model described above is compared with the base Almgren-Chriss model to determine whether it increases/decreases the cost of execution for different types of trades consistently and significantly. This study will help determine whether reinforcement learning is a viable technique which can be used to extend existing closed-form solutions to exploit the nuances of the microstructure where the algorithms are applied.

This paper proceeds as follows: Section 2 introduces the standard Almgren-Chriss model. Section 3 describes the specific hybrid reinforcement learning technique proposed, along with a discussion regarding convergence to optimum action values. Section 4 discusses the data used and results, comparing the 2 models for multiple trade types. Section 5 concludes and proposes some extensions for further research. 


\section{The Almgren-Chriss model}

Bertsimas and Lo are pioneers in the area of optimal liquidation, treating the problem as a stochastic dynamic programming problem \cite{bertsimas}. They employed a dynamic optimisation procedure which finds an explicit closed-form best execution strategy, minimising trading costs over a fixed period of time for large transactions. Almgren and Chriss extended the work of \cite{bertsimas} to allow for risk aversion in their framework \cite{almgren}. They argue that incorporating the uncertainty of execution of an optimal solution is consistent with a trader's utility function. In particular, they employ a price process which permits linear permanent and temporary price impact functions to construct an efficient frontier of optimal execution. They define a trading strategy as being \textit{efficient} if there is no strategy which has lower execution cost variance for the same or lower level of expected execution cost. 

The exposition of their solution is as follows: They assume that the security price evolves according to a discrete arithmetic random walk:
\[ S_k = S_{k-1} + \sigma \tau^{1/2} \xi_k - \tau g(\frac{n_k}{\tau}), \]
where:
\begin{align*}
	&S_k = \text{ price at time $k$},\\
	&\sigma = \text{ volatility of the security},\\
	&\tau = \text{ length of discete time interval},\\
	&\xi_k = \text{ draws from independent random variables,}\\
	%&\text{			each with zero mean and unit variance},\\
	&n_k = \text{ volume traded at time $k$ and}\\
	&g(.) = \text{ permanent price impact}.
\end{align*}

Here, permanent price impact refers to changes in the equilibrium price as a direct function of our trading, which persists for at least the remainder of the liquidation horizon. Temporary price impact refers to adverse deviations as a result of absorbing available liquidity supply, but where the impact dissipates by the next trading period due to the resilience of the order book. Almgren and Chriss introduce a temporary price impact function $h(v)$ to their model, where $h(v)$ causes a temporary adverse move in the share price as a function of our trading rate $v$ \cite{almgren}. Given this addition, the actual security transaction price at time $k$ is given by:
\[ \tilde{S_k} = S_{k-1} - h(\frac{n_k}{\tau}). \]

Assuming a \textit{sell} program, we can then define the total trading revenue as:
\[ \sum\limits_{k=1}^{N} n_k \tilde{S_k} = X S_0 + \sum\limits_{k=1}^{N} (\sigma \tau^{1/2} \xi_k - \tau g(\frac{n_k}{\tau}) ) x_k - \sum\limits_{k=1}^{N} n_k h(\frac{n_k}{\tau}), \]
where $x_k = X - \sum\limits_{j=1}^{k}n_j = \sum\limits_{j=k+1}^{N}n_j$ for $k=0,1,...,N$.

The total cost of trading is thus given by $x = X S_0 - \sum n_k \tilde{S_k}$, i.e. the difference between the target revenue value and the total actual revenue from the execution. This definition refers to Perold's implementation shortfall measure (see \cite{perold}), and serves as the primary transaction cost metric which is minimised in order to maximise trading revenue. Since implementation shortfall is a random variable, Almgren and Chriss compute the following:
\[ E(x) := \sum\limits_{k=1}^{N} \tau x_k g(\frac{n_k}{\tau}) + \sum\limits_{k=1}^{N} n_k h(\frac{n_k}{\tau}) \]
and
\[ Var(x) := \sigma^2 \sum\limits_{k=1}^{N} \tau {x_k}^2. \]
The distribution of implementation shortfall is Gaussian if the $\xi_k$ are Gaussian.

Given the overall goal of minimising execution costs and the variance of execution costs, they specify their objective function as:
\[ \min\limits_{x} \{E(x) + \lambda Var(x) \}, \]
where:
\begin{align*}
	&x = \text{ implementation shortfall}, \\
	&\lambda = \text{ level of risk aversion}.
\end{align*}

The intuition of this objective function can be thought of as follows: Consider a stock which exhibits high price volatility and thus a high risk of price movement away from the reference price. A risk averse trader would prefer to trade a large portion of the volume immediately, causing a (known) price impact, rather than risk trading in small increments at successively adverse prices. Alternatively, if the price is expected to be stable over the liquidation horizon, the trader would rather split the trade into smaller sizes to avoid price impact. This trade-off between speed of execution and risk of price movement is what governs the shape of the resulting trade trajectory in the AC framework.

%Then, assuming that the trade list ($n_j$) does not change sign, the optimal value function becomes:
%\[ U(x) = E(x) + \lambda Var(x), \]
%which is a quadratic function of control parameters $x_1,...,x_{N-1}$ and strictly convex for $\lambda > 0$. The solution to $\frac{\partial U}{\partial x_j} = 0$ yields the following general solution:
A detailed derivation of the general solution can be found in \cite{almgren}. Here, we state the general solution:
\[ x_j = \frac{\sinh ( \kappa ( T-t_j))}{ \sinh (\kappa T)}X \text{ for } j=0,...,N.  \]
The associated trade list is:
\[ n_j = \frac{2 \sinh (\frac{1}{2} \kappa \tau )}{ \sinh (\kappa T) } \cosh (\kappa (T - t_{j-\frac{1}{2}})) X \text{ for } j=0,...,N ,  \]
where:
\begin{align*}
	&\kappa = \frac{1}{\tau}\cosh^{-1}\left(\frac{\tau^2}{2}\tilde{\kappa}^2 + 1\right), \\
	&\tilde{\kappa}^2 = \frac{\lambda \sigma^2}{\eta (1-\frac{\rho \tau}{2 \eta})}, \\
	&\eta = \text{temporary price impact parameter},\\
	&\rho = \text{permanent price impact parameter},\\
	&\tau = \text{length of discrete time period}.
\end{align*}

This implies that for a program of selling an initially long position, the solution decreases monotonically from its initial value to zero at a rate determined by the parameter $\kappa$. If trading intervals are short, $\kappa^2$ is essentially the ratio of the product of volatility and risk-intolerance to the temporary transaction cost parameter. We note here that a larger value of $\kappa$ implies a \textit{more rapid} trading program, again conceptually confirming the propositions of \cite{huberman} that an intolerance for execution risk leads to a larger concentration of quantity traded early in the trading program. Another consequence of this analysis is that different sized baskets of the same securities will be liquidated in the same manner, barring scale differences and provided the risk aversion parameter $\lambda$ is held constant. This may be counter-intuitive, since one would expect larger baskets to be effectively less liquid, and thus follow a \textit{less rapid} trading program to minimise price impact costs.

It should be noted that the AC solution yields a suggested volume trajectory over the liquidation horizon, however there is no discussion in \cite{almgren} as to the prescribed \textit{order type} to execute the trade list. We have assumed that the trade list can be executed as a series of \textit{market orders}. Given that this implies we are always crossing the spread, one needs to consider that traversing an order book with thin volumes and widely-spaced prices could have a significant transaction cost impact. We thus consider a reinforcement learning technique which learns \textit{when} and \textit{how much} to cross the spread, based on the current order book dynamics.

The general solution outlined above assumes linear price impact functions, however the model was extended by Almgren in \cite{almgren2003} to account for non-linear price impact. This extended model can be considered as an alternative base model in future research. 
%The Almgren-Chriss model is widely used and cited in the literature and many sell-side investment banks/algorithm providers use a similar conceptual framework to govern their arrival price benchmark algorithms. It is for this reason that we have used the Almgren-Chriss model as an acceptable standard model, with which we will compare an alternative implementation using reinforcement learning.
\section{A reinforcement learning approach}

The majority of reinforcement learning research is based on a formalism of Markov Decision Processes (MDPs) \cite{barto}. In this context, reinforcement learning is a technique used to numerically solve for a calibrated policy mapping states to optimal or near-optimal actions. It is a framework within which a learning agent repeatedly observes the state of its environment, and then performs a chosen action to service some ultimate goal. Performance of the action has an immediate numeric reward or penalty and changes the state of the environment \cite{dayan}. The problem of solving for an optimal policy mapping states to actions is well-known in stochastic control theory, with a significant contribution by Bellman \cite{bellman}. Bellman showed that the computational burden of an MDP can be significantly reduced using what is now known as dynamic programming. It was however recognised that two significant drawbacks exist for classical dynamic programming: Firstly, it assumes that a complete, known model of the environment exists, which is often not realistically obtainable. Secondly, problems rapidly become computationally intractable as the number of state variables increases, and hence, the size of the state space for which the value function must be computed increases. This problem is referred to as the \textit{curse of dimensionality} \cite{sutton}.

Reinforcement learning offers two advantages over classical dynamic programming: Firstly, agents learn online and continually adapt while performing the given task. Secondly, the methods can employ function approximation algorithms to represent their knowledge. This allows them to generalize across the state space so that the learning time scales much better \cite{dietterich}. Reinforcement learning algorithms do not require knowledge about the exact model governing an MDP and thus can be applied to MDP's where exact methods become infeasible.

Although a number of implementations of reinforcement learning exist, we will focus on \textit{Q-learning}. This is a model-free technique first introduced by \cite{watkins}, which can be used to find the optimal, or near-optimal, action-selection policy for a given MDP.

During \textit{Q-learning}, an agent's learning takes place during sequential episodes. Consider a discrete finite world where at each step $n$, an agent is able to register the current state $x_n\in X$ and can choose from a finite set of actions $a_n\in A$. The agent then receives a probabilistic reward $r_n$, whose mean value $R_{x_n}(a_n)$ depends only on the current state and action. According to \cite{watkins}, the state of the world changes probabilistically to $y_n$ according to:
\[\textit{Prob}(y_n = y | x_n,a_n) = P_{x_n y}(a_n).\]
The agent is then tasked to learn the optimal policy mapping states to actions, i.e. one which maximises total discounted expected reward. Under some policy mapping $\pi$ and discount rate $\gamma$ ($0<\gamma<1$), the value of state $x$ is given by:
\[V^{\pi}(x) = R_x(\pi(x)) + \gamma \sum\limits_{y}^{} P_{xy}(\pi(x)) V^{\pi}(y).\]
According to \cite{dreyfus} and \cite{ross}, the theory of dynamic programming says there is at least one optimal stationary policy $\pi^*$ such that
\[V^*(x) = V^{\pi^*}(x) = \max_a\{  R_x(a) + \gamma \sum\limits_{y}^{} P_{xy}(a) V^{\pi^*}(y)\}.\]
We also define $Q^{\pi}(x,a)$ as the expected discounted reward from choosing action $a$ in state $x$, and then following policy $\pi$ thereafter, i.e.
\[Q^{\pi}(x,a) = R_x(a) + \gamma \sum\limits_{y}^{}P_{xy}(\pi(x)) V^{\pi}(y).\]

The task of the \textit{Q-learning} agent is to determine $V^*$, $\pi^*$ and $Q^{\pi^*}$ where $P_{xy}(a)$ is unknown, using a combination of exploration and exploitation techniques over the given domain. It can be shown that $V^*(x) = \max_a Q^*(x,a)$ and that an optimal policy can be formed such that $\pi^*(x) = a^*$. It thus follows that if the agent can find the optimal Q-values, the optimal action can be inferred for a given state $x$. It is shown in \cite{dayan} that an agent can learn Q-values via experiential learning, which takes place during sequential episodes. In the $n^{th}$ episode, the agent:
\begin{itemize}
	\item observes its current state $x_n$,
	\item selects and performs an action $a_n$,
	\item observes the subsequent state $y_n$ as a result of performing action $a_n$,
	\item receives an immediate reward $r_n$ and
	\item uses a learning factor $\alpha_n$, which decreases gradually over time.
\end{itemize}
$Q$ is updated as follows:
\begin{eqnarray*}
	Q(x_n,a_n)&=&Q(x_n,a_n) + \\
				& &\alpha_n(r_n + \gamma \max_b{Q(x_{n+1},b)} - Q(x_n,a_n)).
\end{eqnarray*}

Provided each state-action pair is visited infinitely often, \cite{dayan} show that $Q$ converges to $Q^*$ for any exploration policy. Singh et al. provide guidance as to specific exploration policies for asymptotic convergence to optimal actions and asymptotic exploitation under the \textit{Q-learning} algorithm, which we incorporate in our analysis \cite{singh}.

\subsection{Implications of finite-horizon MDP}
The above exposition presents an algorithm which guarantees optimal policy convergence of a stationary infinite-horizon MDP. The stationarity assumption, and hence validity of the above result, needs to be questioned when considering a finite-horizon MDP, since states, actions and policies are time-dependent \cite{puterman}.
In particular, we are considering a discrete period, finite trading horizon, which guarantees execution of a given volume of shares. At each decision step in the trading horizon, it is possible to have different state spaces, actions, transition probabilities and reward values. Hence the above model needs revision. Garcia and Ndiaye consider this problem and provide a model specification which suits this purpose \cite{garcia}. They propose a slight modification to the Bellman optimality equations shown above:
\[V_n^*(x) =  \max_{a_n}\{  R_{x}(a_n) + \gamma \sum\limits_{y}^{} P_{xy}^n(a_{n+1}) V_{n+1}^{\pi^*}(y)\}\]
for all $x \in S_n$, $y \in S_{n+1}$, $a_n \in S_n$, $n \in \{0,1,...,N\}$ and $V_{N+1}^*(x) = 0$. This optimality equation has a single solution $V^* = \{V^*_1, V^*_2,...,V^*_N\}$ that can be obtained using dynamic programming techniques. The equivalent discounted expected reward specification thus becomes:
\[Q_n^{\pi}(x,a_n) = R_x(a_n) + \gamma \sum\limits_{y}^{}P^n_{xy}(\pi(x)) V_{n+1}^{\pi}(y).\]
They propose a novel transformation of an $N$-step non-stationary MDP into an infinite-horizon process (\cite{garcia}). This is achieved by adding an artifical final reward-free absorbing state $x_{abs}$, such that all actions $a_N \in A_N$ lead to $x_{abs}$ with probability 1. Hence the revised \textit{Q-learning} update equation becomes:
\[Q_{n+1}(x_n,a_n) = Q_n(x_n,a_n) + \alpha_n(x_n,a_n)U_n,\]
where
\[U_n =
	\begin{cases}
		r_n + \gamma \max_b{Q_n(y_n,b)} - Q_n(x_n,a_n) &\text{ if } x_n \in S_i, i<N\\
		r_n - Q_n(x_n,a_n) & \text{ if } x_n \in S_N \\
		0 & \text{ otherwise. }
	\end{cases}
\]
If $x_n \notin S_N$, $y_n = x_{n+1}$, otherwise choose randomly in $S_1$. If $x_{n+1} \in S_j$, select $a_{n+1} \in A_j$. The learning rule for $S_N$ is thus equivalent to setting $V^*_{N+1}(x_{abs}) = Q^*_{N+1}(x_{abs},a_j) = 0$ $\forall$ $a_j \in A_{N+1}$.

Garcia and Ndiaye further show that the above specification (in the case where $\gamma=1$) will converge to the optimal policy with probability one, provided that each state-action pair is visited infinitely often, $\sum_n\alpha_n(x,a) = \infty$ and $\sum_n^{}\alpha_n^2(x,a) < \infty$ \cite{garcia}.

\subsection{Implementation for optimal liquidation}
Given the above description, we are able to discuss our specific choices for state attributes, actions and rewards in the context of the optimal liquidation problem. We need to consider a specification which adequately accounts for our state of execution and the current state of the limit order book, representing the opportunity set for our ultimate goal of executing a volume of shares over a fixed trading horizon.

\subsubsection{States}
We acknowledge that the complexity of the financial system cannot be distilled into a finite set of states and is not likely to evolve according to a Markov process. However, we conjecture that the essential features of the system can be sufficiently captured with some simplifying assumptions such that meaningful insights can still be inferred. For simplicity, we have chosen a look-up table representation of $Q$. Function approximation variants may be explored in future research for more complex system configurations. As described above, each state $x_n \in X$ represents a vector of observable attributes which describe the configuration of the system at time $n$. As in \cite{nevmyvaka}, we use \textit{Elapsed Time $t$} and \textit{Remaining Inventory $i$} as private attributes which capture our state of execution over a finite liquidation horizon $T$. Since our goal is to modify a given volume trajectory based on favourable market conditions, we include \textit{spread} and \textit{volume} as candidate market attributes. The intuition here is that the agent will learn to increase (decrease) trading activity when \textit{spreads} are narrow (wide) and \textit{volumes} are high (low). This would ensure that a more significant proportion of the total volume-to-trade would be secured at a favourable price and, similarly, less at an unfavourable price, ultimately reducing the post-trade implementation shortfall. Given the look-up table implementation, we have simplified each of the state attributes as follows:
\begin{itemize}
\footnotesize
	\item $T$ = Trading Horizon , $V$ = Total Volume-to-Trade,
	\item $H$ = Hour of day when trading will begin,
	\item $I$ = Number of remaining inventory states,
	\item $B$ = Number of spread states,
	\item $W$ = Number of volume states,
	\item $sp_n$ = \%ile Spread of the $n^{th}$ tuple,
	\item $vp_n$ = \%ile Bid/Ask Volume of the $n^{th}$ tuple,
	\item \textbf{Elapsed Time}: $t_n = 1, 2, 3, ..., T$ ,
	\item \textbf{Remaining Inventory}: $i_n = 1, 2, 3, ..., I$ ,
	\item \textbf{Spread State}: $s_n = \begin{cases} 1, & \mbox{if } 0 < sp_n \le \frac{1}{B} \\ 2, & \mbox{if } \frac{1}{B} < sp_n \le \frac{2}{B} \\ ... \\ B, & \mbox{if } \frac{(B-1)}{R} < sp_n \le 1, \end{cases}$
	\item \textbf{Volume State}: $v_n = \begin{cases} 1, & \mbox{if } 0 < vp_n \le \frac{1}{U} \\ 2, & \mbox{if } \frac{1}{W} < vp_n \le \frac{2}{W} \\ ... \\ W, & \mbox{if } \frac{(W-1)}{W} < vp_n \le W. \end{cases}$
\end{itemize}\normalsize
Thus, for the $n^{th}$ episode, the state attributes can be summarised as the following tuple:
\[ z_n = <t_n, i_n, s_n, v_n>. \]
For $sp_n$ and $vp_n$, we first construct a historical distribution of spreads and volumes based on the training set. It has been empirically observed that major equity markets exhibit \textit{U}-shaped trading intensity throughout the day, i.e. more activity in mornings and closer to closing auction. A further discussion of these insights can be found in \cite{admati} and \cite{brock}. In fact, \cite{dupreez} empirically demonstrates that South African stocks exhibit similar characteristics. We thus consider simlulations where training volume/spread tuples are \textit{H}-hour dependent, such that the optimal policy is further refined with respect to trading time (\textit{H}).

\subsubsection{Actions}
Based on the Almgren-Chriss (AC) model specified above, we calculate the AC volume trajectory ($AC_1, AC_2, ..., AC_T$) for a given volume-to-trade ($V$), fixed time horizon ($T$) and discrete trading periods ($t = 1,2, ..., T$). $AC_t$ represents the proportion of $V$ to trade in period $t$, such that $\sum\limits_{t=1}^{T}AC_t = V$. For the purposes of this study, we assume that each child order is executed as a \textit{market order} based on the prevailing limit order book structure. We would like our learning agent to modify the AC volume trajectory based on prevailing volume and spread characteristics in the market. As such, the possible actions for our agent include:
\begin{itemize}	
	\item $\beta_j$ = Proportion of $AC_t$ to trade,
	\item $\beta_{LB}$ = Lower bound of volume proportion to trade,
	\item $\beta_{UB}$ = Upper bound of volume proportion to trade,	
	\item \textbf{Action}: $a_{jt} = \beta_j AC_t$, where $\beta_{LB} \le \beta_j \le \beta_{UB}$ \\ and $\beta_{j}=\beta_{j-1}+\beta_{incr}.$
\end{itemize}
The aim here is to train the learning agent to trade a higher (lower) proportion of the overall volume when conditions are favourable (unfavourable), whilst still broadly preserving the volume trajectory suggested by the AC model. To ensure that the total volume-to-trade is executed over the given time horizon, we execute any residual volume at the end of the trading period with a \textit{market order}.

\subsubsection{Rewards}
Each of the actions described above results in a volume to execute with a \textit{market order}, based on the prevailing structure of the limit order book. The size of the child order volume will determine how deep we will need to traverse the order book. For example, suppose we have a \textit{BUY} order with a \textit{volume-to-trade} of 20000, split into child orders of 10000 in period $t$ and 10000 in period $t+1$. If the structure of the limit order book at time $t$ is as follows:
\begin{itemize}\footnotesize
	\item \textit{Level-1 Ask Price} = 100.00; \textit{Level-1 Ask Volume} = 3000
	\item \textit{Level-2 Ask Price} = 100.50; \textit{Level-2 Ask Volume} = 4000
	\item \textit{Level-3 Ask Price} = 102.30; \textit{Level-3 Ask Volume} = 5000
	\item \textit{Level-4 Ask Price} = 103.00; \textit{Level-4 Ask Volume} = 6000
	\item \textit{Level-5 Ask Price} = 105.50; \textit{Level-5 Ask Volume} = 2000	
\end{itemize}\normalsize
the volume-weighted execution price will be:\small 
\[\frac{(3000 \times 100) + (4000 \times 100.5) + (3000 \times 102.3)}{10000} = 100.9 .\] \normalsize
Trading more (less) given this limit order book structure will result in a higher (lower) volume-weighted execution price. If the following trading period $t+1$ has the following structure:
\begin{itemize}\footnotesize
	\item \textit{Level-1 Ask Price} = 99.80; \textit{Level-1 Ask Volume} = 6000
	\item \textit{Level-2 Ask Price} = 99.90; \textit{Level-2 Ask Volume} = 2000
	\item \textit{Level-3 Ask Price} = 101.30; \textit{Level-3 Ask Volume} = 7000
	\item \textit{Level-4 Ask Price} = 107.00; \textit{Level-4 Ask Volume} = 3000
	\item \textit{Level-5 Ask Price} = 108.50; \textit{Level-5 Ask Volume} = 1000	
\end{itemize}\normalsize
the volume-weighted execution price for the second child order will be: \small
\[\frac{(6000 \times 99.8) + (2000 \times 99.9) + (2000 \times 101.3)}{10000} = 100.12 .\] \normalsize
If the reference price of the stock at $t=0$ is 99.5, then the \textit{implementation shortfall} from this trade is:\small
\[\frac{((20000 \times 99.5) - (10000 \times 100.9 + 10000 \times 100.12)}{20000 \times 99.5}\]
\[ = -0.0101 = -101 bps.\] \normalsize
Since the conditions of the limit order book were more favourable for \textit{BUY} orders in period $t+1$, if we had modified the child orders to, say 8000 in period $t$ and 12000 in period $t+1$, the resulting \textit{implementation shortfall} would be:\small
\[\frac{((20000 \times 99.5) - (8000 \times 100.54 + 12000 \times 100.32)}{20000 \times 99.5}\]
\[ = -0.0091 = -91 bps.\]\normalsize
In this example, increasing the child order volume when \textit{Ask Prices} are lower and \textit{Level-1 Volumes} are higher decreases the overal cost of the trade. It is for this reason that \textit{implementation shortfall} is a natural candidate for the rewards matrix in our reinforcement learning system. Each action implies a child order volume, which has an associated volume-weighted execution price. The agent will learn the consequences of each action over the trading horizon, with the ultimate goal of minimising the total trade's \textit{implementation shortfall}.

\subsubsection{Algorithm and Methodology}
Given the above specification, we followed the following steps to generate our results:
\begin{itemize}
	\item Specify a stock ($S$), volume-to-trade ($V$), time horizon ($T$), and trading datetime (from which the trading hour $H$ is inferred),
	\item Partition the dataset into independent \textit{training sets} and \textit{testing sets} to generate results (the \textit{training set} always pre-dates the \textit{testing set}),
	\item Calibrate the parameters for the Almgren-Chriss (AC) volume trajectory ($\sigma, \eta$) using the historical \textit{training set}; set $\rho=0$, since we assume order book is resilient to trading activity (see below),
	\item Generate the AC volume trajectory ($AC_1,...,AC_T$),
	\item Train the \textit{Q-matrix} based on the state-action tuples generated by the \textit{training set},
	\item Execute the AC volume trajectory at the specified trading datetime ($H$) on each day in the \textit{testing set}, recording the \textit{implementation shortfall},
	\item Use the trained \textit{Q-matrix} to modify the AC trajectory as we execute $V$ at the specified trading datetime, recording the \textit{implementation shortfall} and
	\item Determine whether the reinforcement learning (RL) model improved/worsened realised \textit{implementation shortfall}.
\end{itemize}
In order to train the \textit{Q-matrix} to learn the optimal policy mapping, we need to traverse the training data set ($T \times I \times A$) times, where $A$ is the total number of possible actions. The following pseudo-code illustrates the algorithm used to train the \textit{Q-matrix}:

\scriptsize\begin{verbatim}
Optimal_strategy<V,T,I,A>
   For (Episode 1 to N) {
      Record reference price at t=0    
      For t = T to 1 {
         For i = 1 to I
            Calculate episode's STATE attributes <s,v>
            For a = 1 to A {
                Set x = <t,i,s,v>
                Determine the action volume a
                Calculate IS from trade, R(x,a)
                Simulate transition x to y
                Look up max_p Q(y,p)
                Update Q(x,a) = Q(x,a) + alpha*U }}}
Select the lowest-IS action max_p Q(y,p) for optimal policy
\end{verbatim}\normalsize
An important assumption in this model specification is that our trading activity does not affect the market attributes. Although temporary price impact is incorporated into execution prices via depth participation of the \textit{market order} in the prevailing limit order book, we assume the limit order book is resilient with respect to our trading activity. Market resiliency can be thought of as the number of quote updates before the market's spread reverts to its competitive level. Degryse et al. showed that a pure limit order book market (Euronext Paris) is fairly resilient with respect to most order sizes, taking on average 50 quote updates for the spread to normalise following the most aggressive orders \cite{degryse}. Since we are using 5-minute trading intervals and small trade sizes, we will assume that any permanent price impact effects dissipate by the next trading period. A preliminary analysis of South African stocks revealed that there were on average over 1000 quote updates during the 5-minute trading intervals and the pre-trade order book equilibrium is restored within 2 minutes for large trades. The validity of this assumption however will be tested in future research, as well as other model specifications explored which incorporate permanent effects in the system configuration.

\section{Data and results}

\subsection{Data used}
For this study, we collected 12 months of market depth tick data (Jan-2012 to Dec-2012) from the Thomson Reuters Tick History (TRTH) database, representing a universe of 166 stocks that make up the South African local benchmark index (ALSI) as at 31-Dec-2012. This includes 5 levels of order book depth (bid/ask prices and volumes) at each tick. The raw data was imported into a MongoDB database and aggregated into 5-minute intervals showing average level prices and volumes, which was used as the basis for the analysis.

\subsection{Stocks, parameters and assumptions}
To test the robustness of the proposed model in the South African (SA) equity market we tested a variety of stock types, trade sizes and model parameters. Due to space constraints, we will only show a representative set of results here that illustrate the insights gained from the analysis. The following summarises the stocks, parameters and assumptions used for the results that follow:
\begin{itemize}
\footnotesize
	\item \textbf{Stocks}
		\begin{itemize}
			\item SBK (Large Cap, Financials)
			\item AGL (Large Cap, Resources)
			\item SAB (Large Cap, Industrials)
		\end{itemize}
	\item \textbf{Model Parameters}
		\begin{itemize}
			\item $\beta_{LB}$: 0, $\beta_{UB}$: 2, $\beta_{incr}$: 0.25
			\item $\lambda$: 0.01, $\tau$: 5-min, $\alpha_0$: 1, $\gamma$: 1
			\item $V$: 100 000, 1000 000
			\item $T$: 4 (20-min), 8 (40-min), 12 (60-min)
			\item $H$: 9, 10, 11, 12, 13, 14, 15, 16
			\item $I, B, W$: 5, 10
			\item Buy/Sell: BUY
		\end{itemize}
	\item \textbf{Assumptions}
		\begin{itemize}
			\item Max volume participation rate in order book: 20\%
			\item Market is resilient to our trading activity
		\end{itemize}
\end{itemize}\normalsize
Note, we set $\gamma=1$ since \cite{garcia} states that this is a necessary condition to ensure convergence to the optimal policy with probability one for a finite-horizon MDP. We also choose an arbitrary value for $\lambda$, although sensitivities to these parameters will be explored in future work. AC parameters are calibrated and \textit{Q-matrix} trained over a 6-month \textit{training set} from 1-Jan-2012 to 30-Jun-2012. The resultant AC and RL trading trajectories are then \textit{executed} on each day at the specified trading time $H$ in the \textit{testing set} from 1-Jul-2012 to 20-Dec-2012. The \textit{implementation shortfall} for both models is calculated and the difference recorded. This allows us to construct a distribution of \textit{implementation shortfall} for each of the AC and RL models, and for all trading hours $H = 9,10,...,16$.
\subsection{Results}
Table 1 shows the average \% improvement in median \textit{implementation shortfall} for the complete set of stocks and parameter values. These results suggest that the model is more effective for shorter trading horizons ($T=4$), with an average improvement of up to 10.3\% over the base AC model. This result may be biased due to the assumption of order book resilience. Indeed, the efficacy of the trained Q-matrix may be less reliable for stocks which exhibit slow order book resilience, since permanent price effects would affect the state space transitions. In future work, we plan to relax this order book resilience assumption and incoporate permanent effects into state transitions.

Figure 1 illustrates the improvement in median post-trade \textit{implementation shortfall} when executing the volume trajectories generated by each of the models, for each of the candidate stocks at the given trading times. In general, the RL model is able to improve (lower) ex-post \textit{implementation shortfall}, however the improvement seems more significant for early morning/late afternoon trading hours. This could be due to the increased trading activity at these times, resulting in more state-action visits in the \textit{training set} to refine the associated Q-matrix values. We also notice more dispersed performance between 10:00 and 11:00. This time period coincides with the UK market open, where global events may drive local trading activity and skew results, particularly since certain SA stocks are dual-listed on the London Stock Exchange (LSE). The improvement in \textit{implementation shortfall} ranges from 15 bps (85.3\%) for trading 1000 000 of SBK between 16:00 and 17:00, to -7 bps (-83.4\%) for trading 100 000 SAB between 16:00 and 17:00. Overall, the RL model is able to improve \textit{implementation shortfall} by 4.8\%.
\begin{table}[ht]	
	\centering
	\captionsetup{font=scriptsize}
	\footnotesize\addtolength{\tabcolsep}{-3pt}
	\begin{tabular}{lll cccccccc r}
		\hline\hline
		\multicolumn{3}{c}{Parameters} & \multicolumn{8}{c}{Trading Time(hour)} & \multicolumn{1}{r}{Average} \\
		\hline
		V & T & I,B,W & 9 & 10 & 11 & 12 & 13 & 14 & 15 & 16 & \\
		\hline
		100000	&	4	&	5	&	23.9	&	-1.4	&	4.7	&	13.4	&	1.8	&	3.3	&	1.8	&	35.1	&	10.3\\
		100000	&	8	&	5	&	25.3	&	4.3	&	8.3	&	2.3	&	1.4	&	9.9	&	-0.6	&	-1.9	&	6.1\\
		100000	&	12	&	5	&	32.7	&	-25.2	&	7.2	&	-2.7	&	-1.5	&	4.6	&	4.5	&	-3.3	&	2.1\\
		1000000	&	4	&	5	&	23.3	&	-1.3	&	4.8	&	9.3	&	1.9	&	3.5	&	1.8	&	35.0	&	9.8\\
		1000000	&	8	&	5	&	28.8	&	5.6	&	8.2	&	1.9	&	1.4	&	9.9	&	-0.3	&	-2.6	&	6.6\\
		1000000	&	12	&	5	&	33.1	&	-25.0	&	7.2	&	-4.0	&	-0.8	&	4.8	&	4.8	&	1.2	&	2.7\\
		100000	&	4	&	10	&	22.9	&	1.3	&	3.0	&	9.7	&	2.7	&	5.8	&	3.5	&	-26.1	&	2.8\\
		100000	&	8	&	10	&	26.0	&	4.3	&	6.7	&	-0.2	&	3.5	&	8.6	&	1.6	&	-3.1	&	5.9\\
		100000	&	12	&	10	&	27.8	&	-21.9	&	7.5	&	-4.1	&	0.6	&	1.8	&	6.2	&	-9.5	&	1.1\\
		1000000	&	4	&	10	&	22.6	&	1.4	&	3.1	&	9.3	&	2.5	&	6.0	&	3.6	&	-26.1	&	2.8\\
		1000000	&	8	&	10	&	26.3	&	5.0	&	7.2	&	-0.5	&	3.3	&	7.0	&	2.3	&	-1.8	&	6.1\\
		1000000	&	12	&	10	&	27.9	&	-24.3	&	8.3	&	-6.9	&	0.5	&	1.8	&	7.5	&	-3.3	&	1.4\\
		\hline\hline
	\end{tabular}
	\caption{Average \% improvement in median \textit{implementation shortfall} for various parameter values, using AC and RL models. Training $H$-dependent.}
	\normalsize
	\label{tab:allResults}
	\vspace{-5mm}
\end{table}

\begin{figure}[ht]
\centering
\includegraphics[width=3.5in]{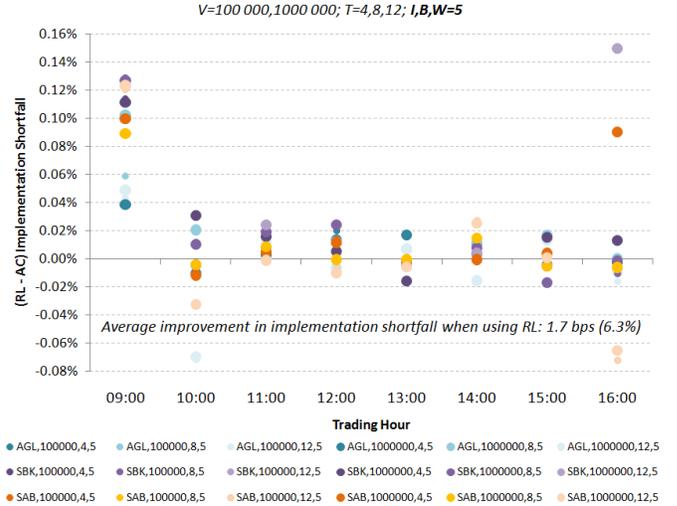}
\captionsetup{font=scriptsize}
\caption{Difference between median implementation shortfall generated using RL and AC models, with given parameters (\textit{I,B,W = 5}). Training $H$-dependent.}
\label{fig_sim}
\vspace{-7mm}
\end{figure}

\begin{figure}[ht]
\centering
\includegraphics[width=3.5in]{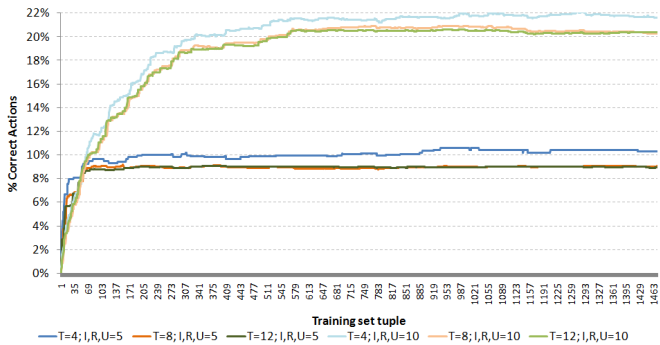}
\captionsetup{font=scriptsize}
\caption{\% correct actions implied by Q-matrix after each training set tuple. Training $H$-dependent.}
\label{fig_sim}
\vspace{-3mm}
\end{figure}

%\begin{figure}[ht]
%\centering
%\includegraphics[width=3.5in]{C://PhD_Work//Papers_and_Conference_Submissions//CIFEr2014//Images//Improvement_IS_H_Figure3.png}
%\caption{Difference between median implementation shortfall generated using RL and AC models, with given parameters (\textit{I,B,W = 10}). Training $H$-dependent.}
%\label{fig_sim}
%\end{figure}

\begin{table}[h]
	\centering
	\footnotesize\addtolength{\tabcolsep}{-3pt}
	\begin{tabular}{ccccc c}
		\hline\hline
		\multicolumn{3}{c}{Parameters} & \multicolumn{2}{c}{Standard Deviation(\%)}& \% improvement\\ % & \multicolumn{2}{c}{IS(bps)} & \% improvement\\
		\hline
		V & T & I,B,W & AC & RL & in IS\\
		\hline
		100000	&	4	&	5	&	0.13	&	0.17	&	10.3\\ %	&	-17.7	&	-15.9	&	10.3\\
		100000	&	8	&	5	&	0.14	&	0.23	&	6.1\\ %	&	-17.6	&	-15.2	&	6.1\\
		100000	&	12	&	5	&	0.14	&	0.26	&	2.1\\ %	&	-17.6	&	-14.7	&	2.1\\
		1000000	&	4	&	5	&	0.13	&	0.17	&	9.8\\ %	&	-17.8	&	-16.0	&	9.8\\
		1000000	&	8	&	5	&	0.14	&	0.23	&	6.6\\ %	&	-17.7	&	-15.3	&	6.6\\
		1000000	&	12	&	5	&	0.14	&	0.26	&	2.7\\ %	&	-17.7	&	-14.8	&	2.7\\
		100000	&	4	&	10	&	0.13	&	0.17	&	2.8\\ %	&	-17.7	&	-16.3	&	2.8\\
		100000	&	8	&	10	&	0.14	&	0.22	&	5.9\\ %	&	-17.6	&	-15.3	&	5.9\\
		100000	&	12	&	10	&	0.14	&	0.26	&	1.1\\ %	&	-17.6	&	-14.8	&	1.1\\
		1000000	&	4	&	10	&	0.13	&	0.17	&	2.8\\ %	&	-17.8	&	-16.3	&	2.8\\
		1000000	&	8	&	10	&	0.14	&	0.22	&	6.1\\ %	&	-17.7	&	-15.3	&	6.1\\
		1000000	&	12	&	10	&	0.14	&	0.26	&	1.4\\ %	&	-17.7	&	-15.0	&	1.4\\
		\hline
		Average & & & 0.14 & 0.22 & 4.8\\ %	&	1	&	1	&	4.8\\
		\hline\hline
	\end{tabular}
	\captionsetup{font=scriptsize}
	\caption{Standard deviation(\%) of implementation shortfall when using AC vs RL models.}	
	\vspace{-7mm}
\end{table}
%\vspace{-5mm}

%\begin{figure}[!t]
%\centering
%\includegraphics[width=3.5in]{C://PhD_Work//Papers_and_Conference_Submissions//CIFEr2014//Images//AllResults_Figure4.png}
%\caption{Average \% improvement in median \textit{implementation shortfall} for various parameter values, using AC and RL models. Training $H$-independent.}
%\label{fig_sim}
%\end{figure}

%\begin{figure}[ht]
%\centering
%\includegraphics[width=3.5in]{C://PhD_Work//Papers_and_Conference_Submissions//CIFEr2014//Images//SABResults_Figure5.png}
%\caption{Average \% improvement in median \textit{implementation shortfall} for SAB (\textit{V=100 000, T=4, I,B,W=10}), using AC and RL models. Effect of \textit{H}-dependent training.}
%\label{fig_sim}
%\end{figure}

%Figures 2 and 5 show the effect of restricting \textit{spread/volume} tuples in the \textit{training set} to those that fall in the same \textit{H}-hour trading time. We notice a significant improvement over the $H$-independent results for SAB, with \textit{T=4, I,B,W=10}. However the results for the other stocks/parameters do not suggest any significant benefit from this additional training constraint. A potential reason could be that the temporal \textit{spread/volume} characteristics of SAB are relatively more persistent and can be exploited to improve the learning agent's performance. Since these effects are not consistent across other stocks, we would need to further investigate which particular features of SAB are driving the improved performance, to determine whether insights can be generalised to other classes of stocks.

Figure 2 shows the \% of \textit{correct actions} implied by the Q-matrix, as it evolves through the training process after each tuple visit. Here, a \textit{correct action} is defined as a reduction (addition) in the volume-to-trade based on the max Q-value action, in the case where \textit{spreads} are above (below) the 50\%ile and \textit{volumes} are below (above) the 50\%ile level. This coincides with the intuitive behaviour we would like the RL agent to learn. These results suggest that finer state granularity ($I,B,W=10$) improves the overall accuracy of the learning agent, as demonstrated by the higher \% \textit{correct actions} achieved. All model configurations seem to converge to some \textit{stationary} accuracy level after approximately 1000 tuple visits, suggesting that a shorter training period may yield similar results. We do however note that improving the \% of correct actions by increasing the granularity of the state space does not necessarily translate into better model performance. This can be seen by Table 1, where the results where $I,B,W = 10$ do not show any significant improvement over those with $I,B,W = 5$. This suggests that the market dynamics may not be fully represented by \textit{volume} and \textit{spread} state attributes, and alternative state attributes should be explored in future work to improve ex-post model efficacy.

Table 2 shows the average standard deviation of the resultant \textit{implementation shortfall} when using each of the AC and RL models. Since we have not explicitly accounted for \textit{variance of execution} in the RL reward function, we see that the resultant trading trajectories generate a higher standard deviation compared to the base AC model. Thus, although the RL model provides a performance improvement over the AC model, this is achieved with a higher degree of execution risk, which may not be acceptable for the trader. We do note that the RL model exhibits comparable risk for $T=4$, thus validating the use of the RL model to reliably improve IS over short trade horizons. A future refinement on the RL model should incorporate \textit{variance of execution}, such that it is consistent with the AC objective function. In this way, a true comparison of the techniques can be done, and one can conclude as to whether the RL model indeed outperforms the AC model at a statistically significant level.

\section{Conclusion}
In this paper, we introduced reinforcement learning as a candidate machine learning technique to \textit{enhance} a given optimal liquidation volume trajectory. Nevmyvaka, Feng and Kearns showed that reinforcement learning delivers promising results where the learning agent is trained to choose the optimal limit order price at which to place the remaining inventory, at discrete periods over a fixed liquidation horizon \cite{nevmyvaka}. Here, we show that reinforcement learning can also be used successfully to modify a given volume trajectory based on market attributes, executed via a sequence of \textit{market orders} based on the prevailing limit order book.

Specifically, we showed that a simple look-up table \textit{Q-learning} technique can be used to train a learning agent to modify a static Almgren-Chriss volume trajectory based on prevailing spread and volume dynamics, assuming order book resiliency. Using a sample of stocks and trade sizes in the South African equity market, we were able to reliably improve post-trade \textit{implementation shortfall} by up to 10.3\% on average for short trade horizons, demonstrating promising potential applications of this technique. Further investigations include incorporating \textit{variance of execution} in the RL reward function, relaxing the order book resiliency assumption and alternative state attributes to govern market dynamics.

%Further investigations into the efficacy of making agent learning trading-hour-dependent would be necessary before substantial conclusions are made, however results for SAB suggest this additional constraint could improve trading performance for certain classes of stocks.

%The analysis presented in this paper can be extended in a number of ways for future research. The following areas present natural extensions:
%\begin{itemize}
%	\item Alternative optimal liquidation base models, including VWAP-benchmark models such as Konishi (2002) \cite{konishi}, McCulloch and Kazakov (2007) \cite{kazakov} and Frei and Westray (2013) \cite{frei}.
%	\item Alternative choices for state attributes, including more complex system configurations which incorporate permanent price impact effects from trading activity. Here we consider the work of Obizhaeva and Wang (2006) \cite{obizhaeva} and Alfonsi et al. (2010) \cite{alfonsi} on modelling order book dynamics and the impact of trading activity.
%\end{itemize} 


\section*{Acknowledgment}
The authors thank Dr Nicholas Westray for his contribution in the initiation of this work, as well as the insightful comments from the anonymous reviewers.
This work is based on the research supported in part by the National Research Foundation of South Africa (Grant Number CPRR 70643)

\end{document}